\documentclass[multphys,vecphys]{svmult}
\usepackage[latin2]{inputenc}
\newcommand{\be}{\begin{equation}}
\newcommand{\ee}{\end{equation}}
\usepackage{amsfonts}
\usepackage{amssymb}
\usepackage{natbib}
\usepackage{makeidx}
\usepackage{graphicx}

\usepackage{multicol}
\usepackage[bottom]{footmisc}

\makeindex

\begin{document}

\title*{Risk portofolio management under Zipf analysis based strategies}
\titlerunning{Risk portofolio management  under Zipf ...}
\author{M. Ausloos  \and  Ph. Bronlet
\institute{SUPRATECS, B5, Sart Tilman Campus, B-4000 Li$\grave
e$ge, Euroland
\newline {\tt marcel.ausloos@ulg.ac.be; philippe.bronlet@ulg.ac.be}}}

\maketitle
\begin{abstract} {A so called Zipf analysis portofolio management 
technique is introduced in order to
comprehend the risk and returns. Two portofoios are built each
from a well known financial index. The
portofolio management is based on two approaches: one called the ''equally
weighted portofolio'', the other the ''confidence parametrized portofolio''.
A discussion of the (yearly) expected return, variance, Sharpe  ratio 
and $\beta$ follows.
Optimization levels of high returns or low risks are found. }

\end{abstract}


\vskip 0.5cm

\section{Introduction}
Risk must be expected for any reasonable investment.
 A portofolio should be constructed such as to minimize
   the investment risk in presence of somewhat unknown
   fluctuation distributions
   of the various asset prices [1,2] in view of
    obtaining the highest possible returns.
   The risk considered hereby is measured through the variances
  of returns, i.e. the $\beta$.
   Our previous approaches were based on the
    ''time dependent'' Hurst exponent [3].
   In contrast, the Zipf method which we previously
   developed as an investment strategy
   (on usual financial indices) [4,5] can be adapted
   to portofolio management.
  This is shown here through portofolios based on
  the $DJIA30$ and the $NASDAQ100$.
  Two strategies are examined through
  different weights to the shares in the portofolio
  at buying or selling time. This is
   shown to have some interesting features.
A key  parameter is the
    {\it coefficient of confidence. }
  Yearly expected levels of returns are
  discussed through the Sharpe ratio and the
  risk through the $\beta$. \nopagebreak

\section{Data}
  Recall that a time series signal can be
  interpreted as a series of words of $m$ letters made of characters
  taken from an alphabet having $k$ letters. Here below
  $k=2$: $u$ and $d$, while the words  have a
  systematic (constant) size ranging between 1 and 10 letters.

    Prior to some strategy definition and implementation, let
   us introduce a few notations. Let the probability of finding a word
   of size $m$ ending with a $u$ in the $i$ (asset) series
   be given  by
   $P_{m,i}(u) \equiv P_i([c_{t-m+2},c_{t-m+1},..., c_{t+1},c_{t};u])$
   and
   correspondingly by $P_{m,i}(d)$
   when a $d$ is the last letter of a word of size $m$.
   The character $c_{t}$ is that seen at the end of day $t$.

  In the following, we have downloaded the daily
   closing price data
   available from the web: (i) for the $DJIA30$, 3909
   data points for the 30 available shares, i.e. for about 16 
years\footnote{From
   Jan. 01, 1989 till Oct. 04, 2004};
   (ii) for the $NASDAQ100$, 3599 data points\footnote{From
   June 27, 1990 till Oct. 04, 2004}
   for the 39 shares which have been maintained in the index, i.e.
   for about 14.5 years.
    The first 2500 days  are taken as
   the preliminary $historical$ data
   necessary for calculating/setting the above probabilities at time $t=0$.
    From these
   we have invented a strategy for the
   following 1408 and 1098 possible investment days, respectively,
   i.e. for ca.
   the latest 6
   and 4.5 years
respectively. The relevant probabilities
are recalculated at the end of each day in order
to implement a $buy$ or $sell$ action on the following day.
The daily
strategy consists in buying a share in any index if
  $P_{m,i}(u)$ $\geq$ $ P_{m,i}(d)$, and in selling it if $P_{m,i}(u)$ 
$\leq$ $P_{m,i}(d)$.

However the weight of a given stock in the portofolio
of $n$ assets can be different according to
the preferred strategy.
In the
  equally
weighted portofolio (EWP), each stock $i$ has the same
weight, i.e. we give
$w_{i\in B} = 2/n_u$  and $w_{i\in S}  = -1/n_d$,
where $n_u$ ($n_d$) is the
number of shares in $B$ ($S$) respectively
such that $\Sigma [ w_{i\in B}  + w_{i\in S} ] =1$, with
$n_u + n_d = n$ of course. This
portofolio management strategy is called $ZEWP$.

In the other strategy, called $ZCPP$, for the confidence
parametrized portofolio (CPP), the weight of a share depends on
a confidence parameter
$K_{m,i}$ $ \equiv $  $P_{m,i}(u)$ -  $P_{m,i}(d)$.
The shares $i$ to be bought on a day belong to the set $B$ when
$K_{m,i}>0$, and those to be sold belong to the set $S$ when
$K_{m,i}<0$.
The respective weights are then taken to be
$w_B = \frac{2 K_{m,i\in B}}{\Sigma K_{m,i \in B}}$, and
$w_S = \frac{- K_{m,i \in S}}{\Sigma K_{m,i \in S}}$.

\section{Results}

The yearly return, variance, Sharpe ratio, and $\beta$
   are given in Table 1 and Table
  2 for the so called $DJIA30$ and so called $NASDAQ39$ shares respectively
  as a function of the word length $m$.
   The last
line gives the corresponding results for the $DJIA30$
and the $NASDAQ100$ respectively. We have calculated the
  average (over 5 or 4 years for the $DJIA30$ and  $NASDAQ39$
  respectively) yearly returns, i.e. $E(r_{P})$ for the portofolio $P$.
  The yearly variances  $\sigma_P$ result from the 5 or 4 years
   root mean square deviations from the mean.
   The Sharpe ratio $SR$ is given by $SR$ = $E(r_{P})$ / $\sigma_P$
   and is considered to measure the portofolio
   performance.
   The $\beta$  is given by $cov(r_P,r_M)/ \sigma^{2}_{M}$
   where the $P$ covariance $cov(r_P,r_M)$ is measured with respect
   to the
   relevant  financial index, so called market ($M$),
    return. Of course, $\sigma^{2}_{M}$ measures the relevant $market$
   variance. The $\beta$  is considered to
   measure the portofolio risk.
  For lack of space the data in the tables are not
  graphically displayed.

It is remarkable that
the $E(r_P)$ is rather low for the $ZEWP$, and so is the
$\sigma_P$,
but the $E(r_{P})$ can be very large, but so is the $\sigma_P$
in the $ZCPP$ case  for both portofolios based on the $DJIA30$.
  The same observation can be made
for the $NASDAQ39$. In the former case, the highest $E(r_{P})$
is larger than 100\% (on average) and
occurs for $m$ =4, but it is the highest for $m$=3 in the latter case.
Yet the risk is large in such cases.
The dependences of the Sharpe ratio and $\beta$  are not smooth
functions of $m$, even indicating some systematic
  dip near $m=6$, in 3 cases; a peak occurs otherwise.

  The expected yearly returns $E(r_{P})$ $vs.$ $\sigma$ are shown for both
  portofolios and for both strategies in Figs.1-2, together
  with the equilibrium line, given by
  $E(r_{M}) (\sigma/\sigma_M)$, where it is understood that
  $\sigma$ is the appropriate value for the investigated case.
Except for rare isolated points below the equilibrium line,
   data points fall above it. They are even very
much above in the $ZCPP$'s.
cases.

\begin{table}[h]
\begin{center}
\begin{tabular}{rrrrrrrrrr}
  \hline
    & \multicolumn{4}{c}{ZEWP} && \multicolumn{4}{c}{ZCPP} \\
  \cline{2-5} \cline{7-10}
  $m$ & $E(r_P) $ & $\sigma_P $ & SR & $\beta$ && $E(r_P) $ & 
$\sigma_P $ & SR & $\beta$ \\
  \hline
  1 &  20.00 & 16.98 & 1.18 &  0.97 &&  20.16 & 17.95 & 1.12 &  1.02 \\
  2 &  18.10 & 16.21 & 1.12 &  0.92 &&  20.36 & 17.66 & 1.15 &  1.00 \\
  3 &  22.00 & 14.05 & 1.57 &  0.79 &&  65.24 & 39.52 & 1.65 &  0.08 \\
  4 &  24.93 & 11.90 & 2.09 &  0.57 && 104.85 & 47.02 & 2.23 & -1.11 \\
  5 &  22.60 &  9.16 & 2.47 &  0.38 &&  95.96 & 56.54 & 1.70 & -1.58 \\
  6 &  18.37 & 11.68 & 1.57 &  0.47 &&  67.97 & 40.55 & 1.68 &  0.09 \\
  7 &  17.33 &  8.93 & 1.94 & -0.06 &&  65.27 & 30.18 & 2.16 & -0.50 \\
  8 &   9.84 &  7.73 & 1.27 &  0.11 &&  53.83 & 37.52 & 1.43 &  0.32 \\
  9 &  11.23 &  4.91 & 2.29 & -0.01 &&  44.23 & 38.12 & 1.16 &  0.58 \\
10 &   6.46 &  7.11 & 0.91 &  0.15 &&  37.40 & 61.05 & 0.61 &  1.92 \\
\hline \hline
     & $E(r_M)$ & $\sigma_M $ & SR & $\beta$\\
\cline{2-5}
DJIA30 & 17.09 & 17.47 & 0.98 & 1 \\
\hline
\end{tabular}
\caption{Statistical results for a portofolio
based on the 30 shares in the  $DJIA30$ index for two strategies, i.e. $ZEWP$
and $ZCPP$
  based on
different word sizes $m$ for the time interval mentioned in the text.
  The last
line gives the corresponding results for the $DJIA30$.
All quantities are given in $\%$}
\label{tab 3.1}
\end{center}
\end{table}

\begin{table}[h]
\begin{center}
\begin{tabular}{rrrrrrrrrr}
  \hline
    & \multicolumn{4}{c}{ZEWP} && \multicolumn{4}{c}{ZCPP} \\
  \cline{2-5} \cline{7-10}
  $m$ & $E(r_P) $ & $\sigma_P $ & SR & $\beta$ && $E(r_P) $ & 
$\sigma_P $ & SR & $\beta$ \\
  \hline
  1 &  12.68 & 22.01 & 0.58 &  0.89 &&   5.41 &  26.30 & 0.21 &  0.55 \\
  2 &  11.43 & 19.99 & 0.57 &  0.81 &&   2.25 &  28.12 & 0.08 &  0.63 \\
  3 &  20.25 & 16.92 & 1.20 &  0.24 && 149.27 & 192.91 & 0.77 & -1.87 \\
  4 &  27.08 & 15.74 & 1.72 & -0.04 && 131.69 & 149.75 & 0.88 & -1.70 \\
  5 &  27.84 & 11.49 & 2.42 & -0.18 && 106.63 & 103.30 & 1.03 & -1.08 \\
  6 &  24.89 &  8.77 & 2.84 & -0.05 &&  90.11 &  68.89 & 1.31 & -0.26 \\
  7 &  15.99 &  9.19 & 1.74 & -0.10 &&  67.28 &  32.58 & 2.07 &  0.37 \\
  8 &  13.93 & 12.39 & 1.13 & -0.25 &&  68.34 &  44.33 & 1.54 &  0.06 \\
  9 &  17.52 & 11.13 & 1.57 & -0.32 &&  99.20 &  38.84 & 2.55 &  0.21 \\
10 &  14.77 & 10.81 & 1.37 & -0.32 &&  71.42 &  32.09 & 2.23 &  0.30 \\
\hline \hline
     & $E(r_M)  $ & $\sigma_M $ & SR & $\beta$\\
\cline{2-5}
NASDAQ100& 7.36 & 24.11 & 0.31 & 1 \\
\hline
\end{tabular}
\caption{Statistical results for a portofolio based on  39 shares 
from the  $NASDAQ100$ index for two strategies,
i.e. $ZEWP$ and $ZCPP$ based on different word sizes $m$ for the time 
interval mentioned in the
text. The last line gives the corresponding results for the $NASDAQ100$.
All quantities are given in $\%$}
\label{tab 3.2}
\end{center}
\end{table}

\begin{figure}
\begin{center}
     \includegraphics[scale=0.4,angle=0]{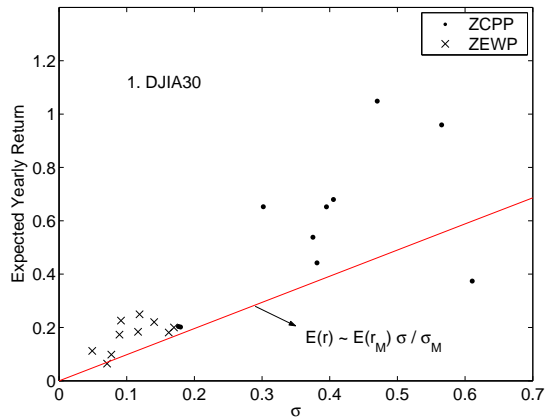}
\end{center}
     \caption{Expected yearly return as a function of the 
corresponding  variance
     for two investment strategies involving the shares in the $DJIA30$.
      The time of investigations concerns the latest 5 yrs }
     \label{fig:1}

\end{figure}

\begin{figure}
\begin{center}
     \includegraphics[scale=0.4,angle=0]{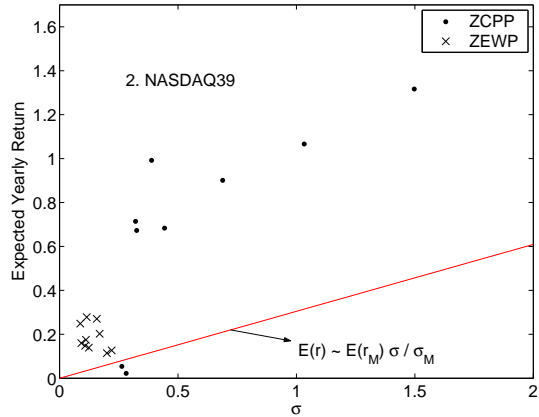}
\end{center}
     \caption{Expected yearly return as a function of the corresponding variance
     for two investment strategies involving 39 shares taken from  the 
$NASDAQ100$.
      The time of investigations concerns the latest 4 yrs}
     \label{fig:2}

\end{figure}

\section{Conclusion}

We have translated the time series of the closing price of stocks
from two financial indices into
letters taken from a two
character alphabet, searched for
   words of $m$ letters, and investigated
the occurrence of such words.
  We have invented two portofolios and maintained them for  a few years,
  buying or selling shares according to
  different strategies. We have calculated
  the corresponding yearly expected return, variance,
Sharpe  ratio and $\beta$. The best returns and
weakest risks have been determined depending on the
  word length. Even though some
risks can be large,  returns are sometimes very high.

\newpage

{\bf Acknowledgments}

\vskip 0.6cm

MA thanks the organizers of the 3rd Nikkei symposium for  financial 
support received
in order to present the above results. \vskip 1cm
\printindex

[1] H.M. Markowitz, Portofolio Selection, J. Finance 8 (1952) 77 - 91.

[2] M. H. Cohen and V.D. Natoli,
  Risk and utility in portfolio optimization, Physica A 324 (2003) 81 - 88.

[3] M. Ausloos, N. Vandewalle and K. Ivanova, Time is Money, in {\it
Noise, Oscillators and Algebraic Randomness}, M. Planat,
  Ed. (Springer, Berlin, 2000) pp. 156-171.

[4]  M. Ausloos and Ph. Bronlet, Strategy for
  Investments from Zipf Law(s),
   Physica A 324 (2003)  30 - 37.

[5] Ph. Bronlet and M. Ausloos, Generalized (m,k)-Zipf law for 
fractional Brownian
  motion-like time series with or without effect of an additional
   linear trend,
  Int. J. Mod. Phys. C. 14 (2003) 351 - 365.

\end{document}